\documentclass[draft]{agujournal}
\draftfalse

\usepackage[utf8]{inputenc}
\usepackage{amssymb}

\newcommand*\aj{AJ}

\newcommand*\apj{ApJ}

\newcommand*\grl{Geophys.~Res.~Lett.}

\newcommand*\icarus{Icarus}

\newcommand*\jgr{J.~Geophys.~Res.}

\newcommand*\nat{Nature}

\newcommand*\planss{Planet.~Space~Sci.}

\newcommand*\ssr{Space~Sci.~Rev.}

\journalname{AGU Books}

\begin{document}

\title{The radiation belts of Jupiter and Saturn}

\authors{E. Roussos\affil{1}, P. Kollmann\affil{2}}

\affiliation{1}{Max Planck Institute for Solar System Research, 37077, Goettingen, Germany}
\affiliation{2}{Johns Hopkins University Applied Physics Laboratory, Laurel, MD 20723-6099, USA}

\correspondingauthor{Elias Roussos}{roussos@mps.mpg.de}

\begin{abstract}

The era of outer planet orbiters (Galileo, Juno and Cassini) is advancing our understanding of how the radiation belts of Jupiter and Saturn are structured, form and evolve well beyond what had been possible during the age of flyby missions and ground-based observations. The nearly two decades-long datasets of these missions, in the context of detailed and long-term observations of Earth's radiation belts, highlight which of the processes that accelerate particles to relativistic kinetic energies and limit their flux intensity can be considered more universal, and thus key for most extraterrestrial magnetospheres, and which reflect the unique aspects of each planet and its magnetospheric system. In this chapter we focus on the in-situ radiation belt observations in the context of theory, simulations and relevant measurements by Earth-based observatories. We describe both the average state and the time variations of Jupiter's and Saturn's radiation belts and associate them with specific physical processes.
\end{abstract}

\section{Introduction}

Similar to Earth, Jupiter and Saturn possess strong internal magnetic fields \citep{connerney2018grl, Dougherty2018science} which react to the incoming solar wind flow and create large magnetospheres. These magnetospheres contain extended, stable, quasi-dipolar regions, one of the basic requirements for an efficient trapping and accumulation of high energy charged particles and thus the formation of radiation belts. On the other hand, Jupiter, Saturn and Earth differ in many other parameters.

In addition to being strongly magnetized, Jupiter and Saturn are fast rotating planets and with powerful plasma sources from active moons embedded within their magnetospheric boundaries. These three factors result in magnetospheres with absolute standoff magnetopause distances that are $\approx$60-99 and $\approx$20-24 times larger than Earth's, for Jupiter and Saturn respectively \citep{joy2002jgr, Kanani2010jgr}. Magnetospheric convection is mostly internally rather than solar wind driven \citep{kivelsonbook}. The two planets' energetic particle trapping regions are very extended (Sections \ref{sec:Saturn}, \ref{sec:Jupiter}) and, especially at low L-shells and altitudes, negligibly disturbed by magnetospheric currents \citep{Birmingham1982jgr}. Furthermore, Jupiter's and Saturn's magnetic moment is oriented northward, resulting in energetic particle gradient and curvature drifts which are reverse in direction compared to those at Earth. Magnetic gradient and curvature electron drifts are opposite to the two planets' rotation and cancel out with the fast corotation drift at relativistic (MeV) energies which are typical for radiation belts. At Earth this happens for much lower energy keV ions \citep{Roederer1970book, Thomsen1980}. The moons and rings contained within the magnetospheres can have a dual role: on the one hand they can supply the magnetosphere with plasma of non-ionospheric or solar wind origin, which not only ``inflates'' them, but also drives waves  that may accelerate particles to radiation belt energies (e.g. \cite{Shprits2018nat}). On the other hand, moons and rings can absorb particles and act simultaneously as particle sinks.

A key question is whether we can consider Earth's radiation belts as a prototype also for Jupiter and Saturn, or if the aforementioned differing aspects of those two planets' magnetospheres have a strong influence on how their radiation belts are structured and evolve. E.g. is the fast corotation of a magnetosphere important just for plasma or also for radiation belt particles? How strongly are the radiation belts of Jupiter and Saturn coupled to their host planets' unique characteristics, such as their moons and rings? To what extent can we probe the geophysics of these systems with radiation belt measurements?

A significant progress to answering such questions became possible only after Jupiter and Saturn were visited by dedicated orbiters (Galileo and Juno at Jupiter, Cassini at Saturn), even though the presence of radiation belts at the two planets was known ever since the detection of synchrotron emissions by Jupiter in 1959 \citep{Drake1959aj} and the short duration Voyager and Pioneer flybys before 1981 (\cite{Sicard2011ieee, Krupp2016} and references therein). The short review that follows highlights observations by these three missions, and in particular in-situ measurements by several energetic particle instruments: the Energetic Particle Detector (EPD) and the Heavy Ion Counter (HIC) on Galileo \citep{Williams1992ssr, garrard1992ssr}, the Magnetosphere Imaging Instrument (MIMI) on Cassini \citep{Krimigis2004ssr} and the Jupiter Energetic-particle Detector Instrument (JEDI) of Juno \citep{Mauk2013ssr}. The long term and continuous measurements of each mission with the same set of instruments mitigates calibration and time variability issues, a typical problem when comparing flyby datasets. Context from plasma, magnetic field and wave measurements (e.g. \citep{Bagenal2016jgr, Persoon2013, Meeks2016, Menietti2016J}), Earth-based observations (e.g. \cite{Tsuchiya2011jgr, Yoshikawa2014ssr}) and simulations (e.g. \cite{Santoscosta2003grl, Shprits2012jgr, Nenon2017jgr}) are key for interpreting in-situ energetic particle measurements and are also discussed, although in less detail.

We start by listing a series of physical processes that are known to operate in Jupiter's and Saturn's radiation belts. Then, by describing each of the two belts' average state and temporal variations, we identify which of the previously listed processes are most important. Ion and electron belts have strong differences in both planets, and thus are discussed separately. In terms of energy, we distinguish the particle populations in two broad ranges above and below $\sim$1 MeV. References to energetic particles outside the radiation belt boundaries are also included, as these can be the belts' seed population. Discussions of time variability concern the radiation belts as a whole. E.g. a galilean satellite may be exposed to varying radiation belt environments, not necessarily because the belts change with time but because the tilted jovian magnetic field wobbles with respect to the moon's orbit. Such variations, which can be important for space weathering of planetary moons (e.g. \citep{milillo2016pss}), are not covered here.

\section{Physical processes}

\subsection{Particle sources and acceleration}\label{sec:sources}

\subsubsection{Plasma sources} 

Radiation belt particles (with the exception of particles from CRAND - see Sec. \ref{sec:CRAND}) start as low energy plasma that is stepwise accelerated to radiation belt energies (Sec. \ref{sec:transport}-\ref{sec:accel}). A major plasma source at the Giant Planets is $SO_2$ and $H_2O$ that originates from the active moons Io \citep{Delamere2003, Kim2018}, Europa \citep{Smyth2006, smith2019apj}, and Enceladus \citep{Cassidy2010icarus, diFabio2011, Blanc2015, Wilson2017, Smith2018} and is in the following ionized and stepwise accelerated through its circulation in the magnetosphere (e.g. \cite{Kivelson2006science, Rymer2008jgr}). Other plasma sources are the inactive moons for example via sputtering and radiolysis \citep{Jurac2001pss, Teolis2010science}, Titan’s atmosphere, through thermal and hydrodynamic escape \citep{Johnson2010book}, the rings \citep{Christon2013, Elrod2014} via photolysis \citep{Johnson2006icarus}, the planetary atmosphere through electron impact dissociation \citep{Tseng2013A}, and the solar wind \citep{diFabio2011, AllenR2018} that can enter through reconnection and Kelvin-Helmholtz instabilities at the magnetopause \citep{Badman2007, Delamere2013jgr}.

\subsubsection{CRAND}\label{sec:CRAND}

CRAND (Cosmic Ray Albedo Neutron Decay) is a direct source of energetic protons and electrons for planetary radiation belts. Galactic cosmic rays (GCR), which are mostly protons at GeV energies \citep{Shikaze2007} that are energetic enough to avoid deflection by the planetary magnetic field \citep{Sauer1980grl, Kotova2018}, can impact a planet's atmosphere or rings. The subsequent nuclear reactions create secondary neutrons \citep{Hess1959physrev, Cooper2018jgr} that decay back into protons at MeV to GeV energies, which can be magnetically trapped and populate the radiation belts. GCR protons impacting a hydrogen atmosphere, as that of Jupiter and Saturn, can also yield neutrons even though the hydrogen nucleus does not include them \citep{Glass1977}. Neutron decay also produces $<1$MeV electrons \citep{Li2017nature} that receive energy both from the decay and the relativistic speed of the parent neutron. Only if the neutron spectrum has high fluxes at GeV energies, $\beta$-decay electron spectra may also reach $>$MeV energies.

Charged GCR secondaries (protons, alpha particles, or fragments of the nucleus) may become trapped on field lines connected to the region on which they originated and are thus expected to return to this region within one bounce period. Such secondaries can therefore only populate a radiation belt if the source material is tenuous enough to make follow-up losses inefficient \citep{roussos2018grl}.

\subsubsection{Radial Transport and adiabatic heating}\label{sec:transport}

Diffusive or convective transport of electrons or ions into stronger magnetic fields leads to their adiabatic acceleration \citep{Roederer1970book}. Diffusive transport across small spatial scales requires processes that move particles randomly both inward and outward \citep{Walt1994book}, such as ultra-low frequency (ULF) electric and magnetic field fluctuations (e.g. \cite{khurana1989jgr}), variable magnetospheric convective flows (e.g. \cite{murakami2016grl}), or changes of thermospheric winds \citep{Brice1973icarus}. Many individual studies provide strong evidence of diffusion acting in both magnetospheres \citep{VanAllen1980jgr, Carbary1983jgr, depater1994jgr, Roussos2007jgr, Tsuchiya2011jgr,  Kita2015jgr, Kollmann2017nature}.

Convective transport occurs on both global and local magnetospheric scales. Global scale convection exists in both magnetospheres \citep{Barbosa1983grl, Ip1983nature, Andriopoulou2012icarus, Thomsen2012jgr, Wilson2013jgr}. On smaller scales, transport is effected through magnetospheric interchange (sometimes referred simply as ``injections''). In such injections, charged particles get transported within few hours across many L-shells along flow channels that are few degrees wide in longitude \citep{Mauk1999jgr, Chen2008jgr, Liu2012, Dumont2014jgr, Paranicas2016icarus}. Interchange is only efficient for $<1$MeV particles, as higher energies drift quickly out of their flow channel \citep{Paranicas2016icarus} and no injection signatures have been observed at higher energies \citep{Paranicas2010jgr, Clark2016jgr}. Injections occur due to a centrifugal interchange instability that occurs at both Gas Giant planets due to their strong plasma sources and centrifugal force onn which this plasma is exposed to ( \cite{Southwood1987, Kivelson2006science}). Interchange, which aims to even out flux tube content decreasing with increasing distance to the planet \citep{Southwood1987}, cannot be described well through diffusion, which smooths out any gradient in energetic particle phase space density \citep{Schulz1974book}. 

\subsubsection{Non-adiabatic acceleration}\label{sec:accel}

Particles can be accelerated locally through field fluctuations with time scales on the order of the gyration time, which drive random energy losses or gains that can be described by energy diffusion \citep{Schulz1974book, Glauert2005, Woodfield2013}. A variety of processes can drive local acceleration, for example whistler mode chorus waves for electrons up to MeV \citep{Woodfield2014, Menietti2014, Soria2017}. The energy for accelerating particles will either be taken from the waves or from another particle population, for example at different pitch angle \citep{Horne2003}. Turbulence near reconnection sites can accelerate various ion species up to hundreds of keV \citep{Radioti2007}. 

The auroral region is also able to accelerate particles to hundreds of keV and MeV energies \citep{McKibben1993pss, roussos2016icarus_a, Palmaerts2016icarus, Clark2017C, Paranicas2018E, Mauk2018, Clark2018}. However, these particles are field-aligned, often to the extent that they will be absorbed in the opposite hemisphere. If instead they scatter and become trapped (e.g. \cite{Speiser1965, Young2008, roussos2016icarus_a}) they can be further accelerated by radial inward transport.

\subsection{Particle sinks and deceleration}\label{sec:sinks}

\subsubsection{Moon absorption}\label{sec:moons}

The Giant Planets have moons orbiting within the radiation belts which absorb energetic particles hitting their surface. To first order, the lifetime of a particle against moon absorption equals the successive encounter time between the moon and the particle and can be very large for electrons that drift with the same speed as a moon orbits around a planet. These are electrons in the MeV range for Saturn and tens of MeV for Jupiter. The encounter probability is reduced if the paths of the moon and particle are not matching due eccentric moon orbits, non-circular particle drifts, or tilts between magnetic and orbit plane. In addition, there is a probability that the charged particle would evade moon absorption due to its gyration, bounce and drift motions, or due to its deflection at field distortions in the local moon-magnetosphere interaction region. For all these reasons, moon absorption probabilities have strong dependencies from energy, species and charged state \citep{Thomsen1980, Hood1983jgr, santoscosta2001pss, Selesnick2009jgr, Krupp2012icarus}.

\subsubsection{Charge exchange}\label{sec:cex}

Singly charged ions $I^+$ can undergo charge exchange with another particle $X$: $I^+ + X^x \rightarrow I + X^{x+1}$. The cross sections for charge exchange are usually larger for neutral particles ($x=0$) than for positive ions \citep{Fujiwara1976}. The energetic ion $I^+$ turns into an energetic neutral atom $I$ (ENA) that is not trapped in the magnetic field anymore and usually escapes the magnetosphere. Charge exchange is therefore a loss process for singly charged ions. In rare cases, the ENA may re-ionize before leaving the magnetosphere, for example when grazing the planet’s atmosphere, and become a source for low altitude particle radiation \citep{Krimigis2005science}. Multiply charged ions \citep{Clark2016jgr, Selesnick2009jgr} can also be affected by charge exchange \citep{Fujiwara1976}, but the process only changes their charge state (that many radiation instruments do not distinguish) while keeping the resulting particle trapped.

\subsubsection{Energy loss}\label{sec:friction}

Energy loss can become equivalent to a particle loss in the context of measurements at a fixed energy range, as it is typical for charged particle detectors \citep{Schulz1974book}. Still, when there are many particles at high energies (hard or rising energy spectrum), these can supply the lower energies, turning the energy loss process at some energies into a source for some others. There are several means how a charged particle can lose energy at Jupiter and Saturn: energy loss due to ionization while it passes through neutral gas or ring grains \citep{Kollmann2013icarus, Kollmann2015jgr}, while it passes through plasma due to collisions with free electrons \citep{Nenon2018}, or while being deflected in strong magnetic fields due to synchrotron emission \citep{santoscosta2001pss}. 

\subsubsection{Pitch angle scattering and friction} \label{sec:scatter}

Various processes may lead trapped particles into the loss cone, reaching into the increasingly dense layers of the exosphere and atmosphere of a planet, where they are absorbed. The change in pitch angle can occur continuously (pitch angle friction) or randomly (scattering). Friction occurs along with synchrotron emission \citep{santoscosta2008pss}.

Scattering has many origins and is usually described through pitch angle diffusion \citep{Schulz1974book, Woodfield2013}. Drivers are wave-particle interactions, for example whistler mode hiss \citep{Glauert2014a, Nenon2017jgr} and chorus \citep{Shprits2012jgr, Soria2017} for electrons and electromagnetic ion cyclotron (EMIC) waves for ions \citep{Nenon2018}. Such waves can arise from particle anisotropies \citep{Kennel1966jgr}, as they can result from radial transport \citep{Bolton1997} or ion pickup \citep{Huddleston1998}.
High radiation intensities can also drive the growth of such waves so that radiation intensities are therefore self-limited to the Kennel-Petschek limit \citep{Kennel1966jgr, Summers2011, Mauk2014}, at least in the absence of strong sources. 

Pitch angle diffusion can also result from scattering in matter \citep{santoscosta2001pss}. Such scattering is most important for electrons or $\lesssim 0.5$MeV ions. Heavy and/or energetic ions do not scatter much even when most of their energy is lost \citep{Kollmann2013icarus, Nenon2018grl}.

\begin{figure}[t]
\includegraphics[clip=true,trim=0cm 0cm 0cm 0cm,width=13cm]{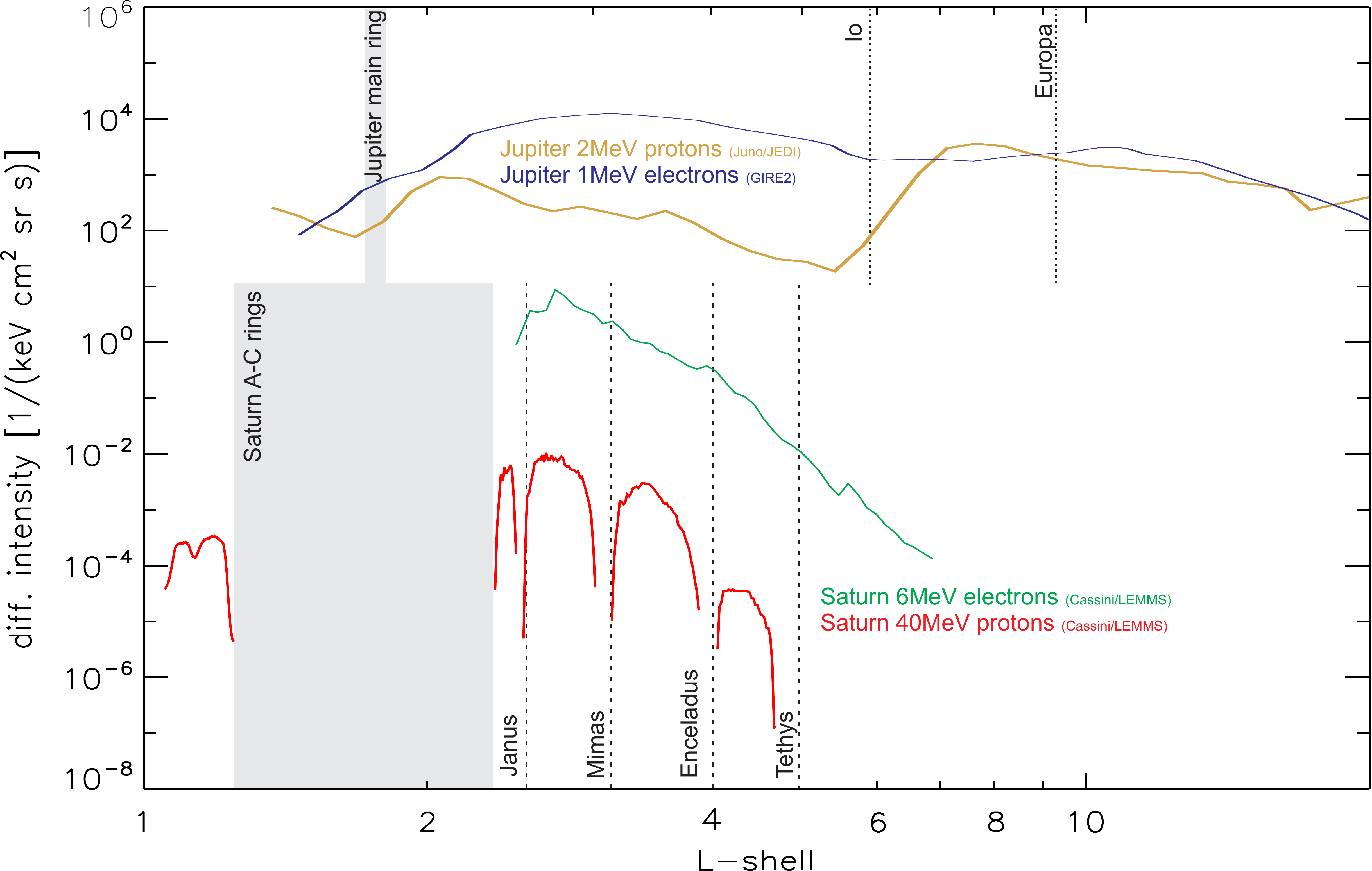}
  \caption{Sample intensity profiles of the radiation belts of Jupiter and Saturn as a function of equatorial distance. Saturn measurements outside of rings: average over Cassini mission \citep{kollmann2018jgr}, Saturn protons inward of rings: corrected data \citep{Roussos2018science}, Jupiter protons: measurements from 2018 DOY 38 (Juno), Jupiter electrons: GIRE2 data compilation \citep{Soria2016}. A minor proton belt component slightly outwards of Saturn's main rings \citep{buratti2018science} is not visible on this scale.}
  \label{fig:radial}
\end{figure}

\section{Saturn's radiation belts}\label{sec:Saturn}

\subsection{Average configuration} 

\subsubsection{Ions}
Saturn’s permanent ion radiation belt (for energies $>10$MeV, where MeV measurements are most reliable) extends from close to Saturn’s visible surface ($L\approx 1.03$) until the orbit of the moon Tethys ($L=4.9$) \citep{VanAllen1980jgrb,Kollmann2013icarus, Roussos2018science}. This belt is segmented into six sectors bound between the planet, the dense (A-C) rings and the orbits of the moons Prometheus/Pandora, Janus/Epimetheus, Mimas, Enceladus and Tethys (Fig. \ref{fig:radial}). Each sector is isolated from the other through moon proton absorption, while Tethys restricts ion transport between the belts and the magnetosphere \citep{Roussos2008grl}. This restriction is the reason protons dominate the ion composition at these energies and L-shells, even though the most abundant ion species in the rest of magnetosphere and below 1 MeV are $H_{0-3}O^+$ \citep{Sergis2007jgr, Wilson2017}. Upper limit intensities of heavier ions in the $>10$MeV/nuc energy range are at least an order of magnitude lower than those of protons \citep{Armstrong2009pss}.

\begin{figure}[t]
$\begin{array}{cc}
\includegraphics[clip=true,trim=0cm 0cm 0cm 0cm,width=6.25cm]{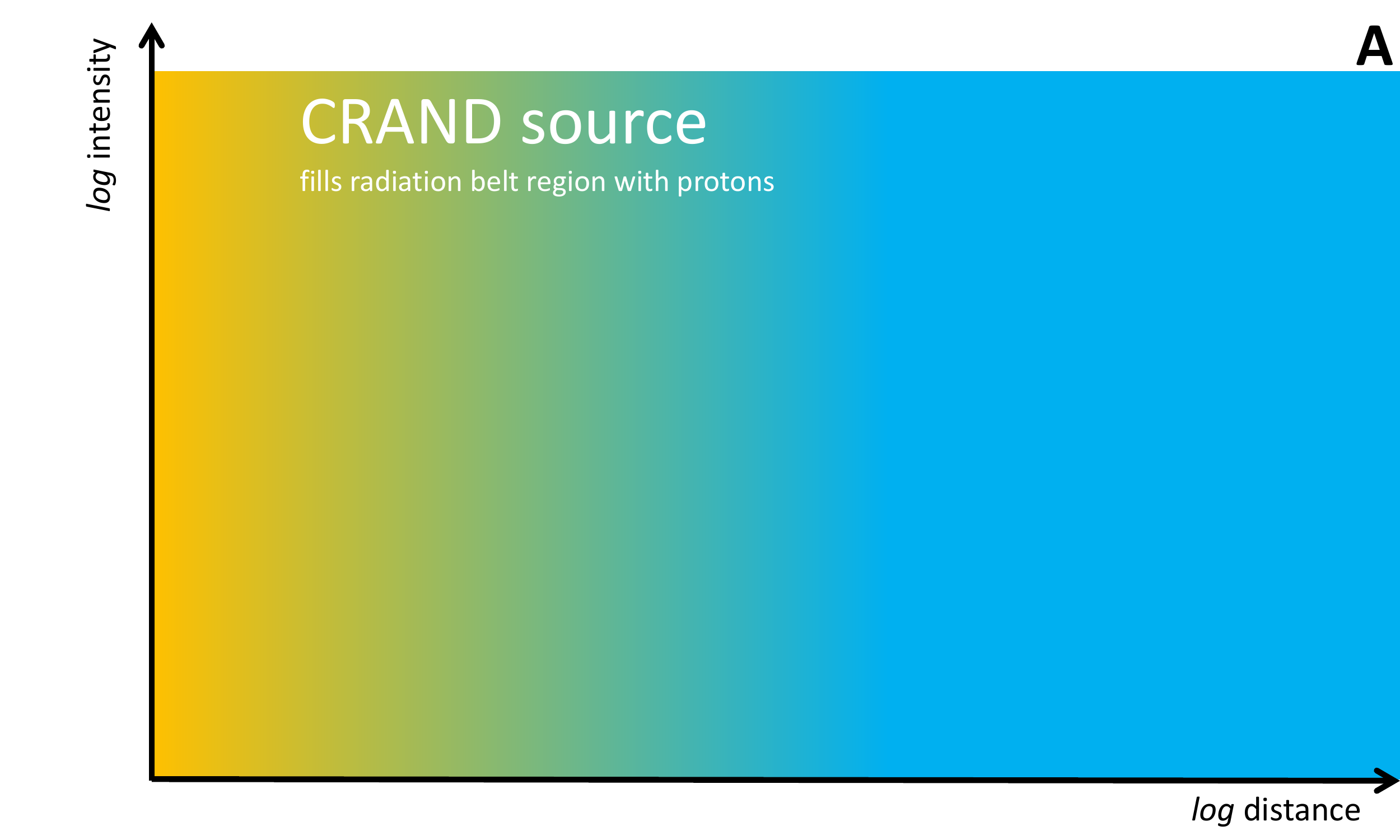} &
\includegraphics[clip=true,trim=0cm 0cm 0cm 0cm,width=6.25cm]{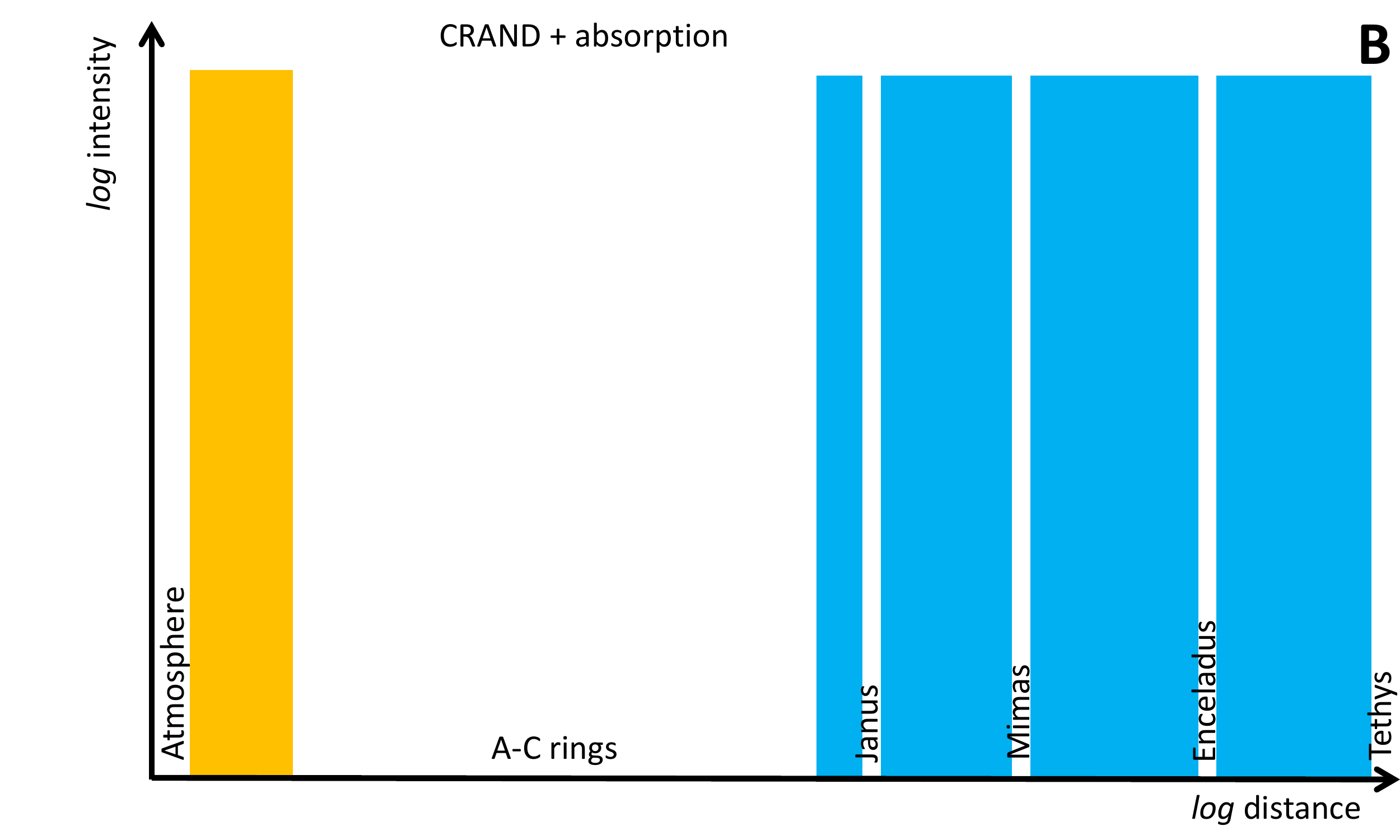} \\
\includegraphics[clip=true,trim=0cm 0cm 0cm 0cm,width=6.25cm]{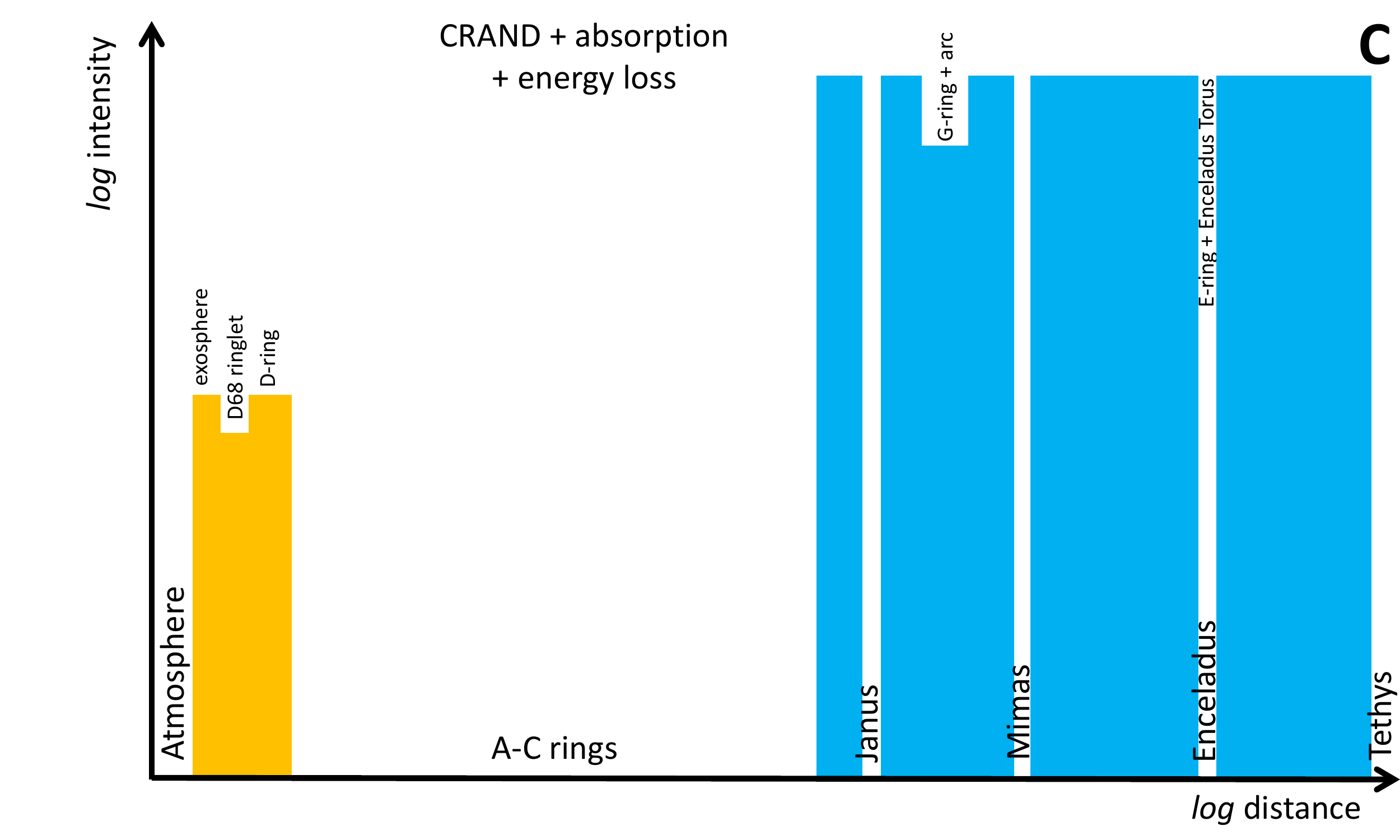} &
\includegraphics[clip=true,trim=0cm 0cm 0cm 0cm,width=6.25cm]{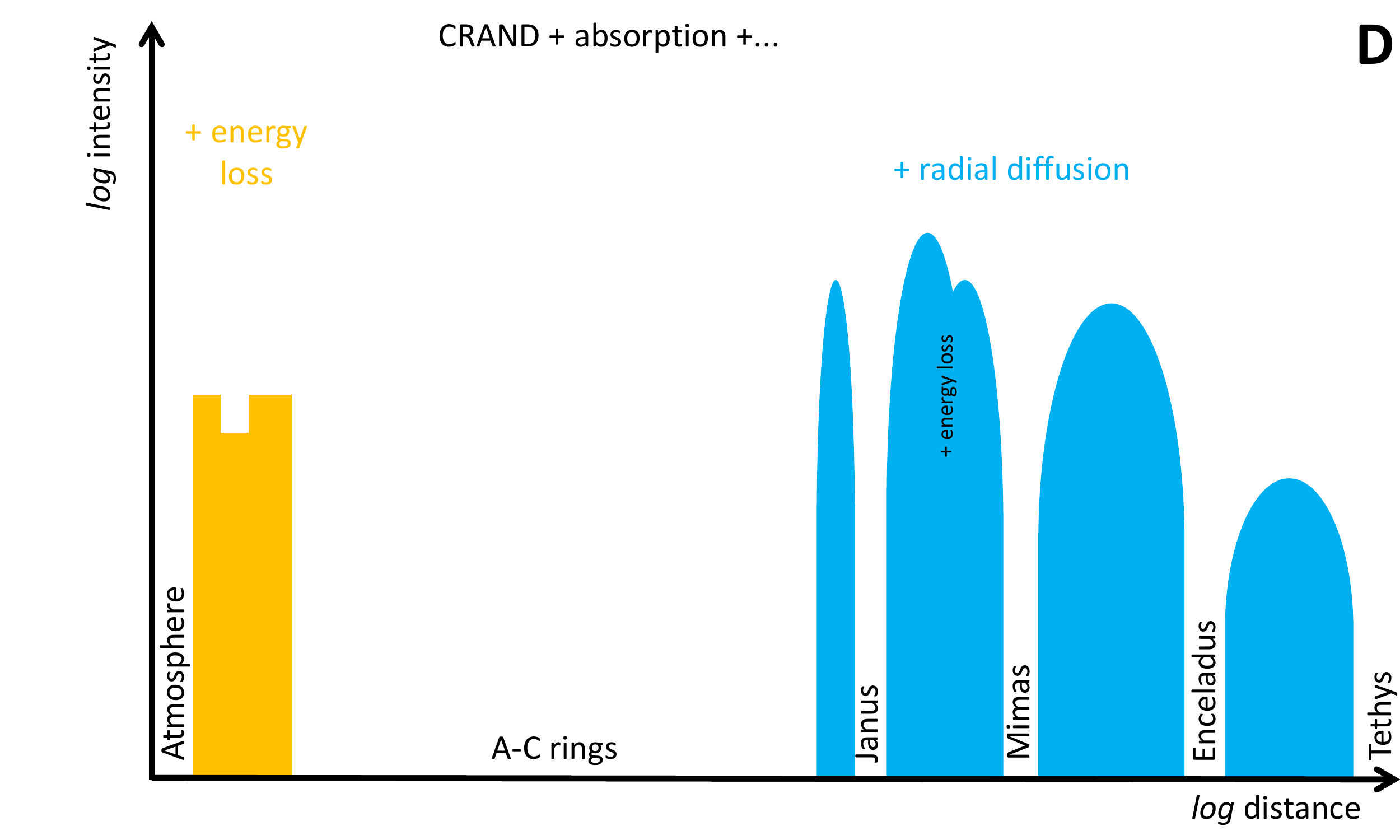} 
\end{array}$
  \caption{Sketch on how different processes shape the radial distribution of Saturn’s proton belts (see red curve in Fig. \ref{fig:radial}). Panel A: CRAND provides the protons (Sec. \ref {sec:CRAND}). Panel B: Moons and dense rings are responsible for the separation of the belts (Sec. \ref {sec:moons}, \ref {sec:friction}). Panel C: Energy loss and absorption in tenuous material changes the shape of the belts (Sec. \ref {sec:friction}). Panel D: Radial diffusion smooths out the edges of the belts (Sec. \ref {sec:transport}).}
  \label{fig:sketch}
\end{figure}

The primary source of these belts is CRAND. CRAND protons may come from both Saturn’s rings and atmosphere for L$>$2.27. Inward of the rings (L$<$1.22) ring CRAND definately dominates (\cite{Cooper1983jgr, Roussos2018science}, Sec. \ref{sec:CRAND}, Fig. \ref{fig:sketch}A). While moons and dense rings absorb CRAND protons along their orbits (Sec. \ref{sec:moons}, Fig. \ref{fig:sketch}B), tenuous material from Saturn’s exosphere and its D- and G- rings reduce the proton intensities only partially (\cite{Roussos2018science, Kollmann2018grl}, Fig. \ref{fig:sketch}C). The Enceladus generated E-ring and gas torus do not have a strong impact on the proton intensities \citep{Kollmann2013icarus}. 

For L$>$2.27 radial diffusion (Sec. \ref{sec:transport}) acts as an additional loss mechanism, since it slowly drives CRAND protons into the moon orbits, regulating the rate at which they get absorbed. This process is also responsible for a broadening of proton depleted regions outside of the moon L-shells. The radial diffusion coefficient increases with distance, causing enhanced losses and lower proton intensities in the outer radiation belts (\cite{Kollmann2013icarus}, Fig. \ref{fig:sketch}D). 

Several ion belt components exist also in the keV energy range. At low altitudes (L$<1.06$), stripped ENAs (Sec. \ref{sec:cex}) are a source of a weak intensity belt comprising tens of keV ions \citep{Krimigis2005science, Roussos2018science, Krupp2018grl}. Outside of the MeV proton belts (L$>$4.9), keV energetic ion fluxes peak at $6\lesssim L\lesssim 9$, depending on energy \citep{Kollmann2011jgr}. The intensities decrease from the peak toward L$\approx 5$, potentially due to charge exchange with the Enceladus neutral torus \citep{Paranicas2008icarus, Dialynas2009jgr}. This ion population rarely penetrates inward of L=4 \citep{roussos2017icarus} and has no sharp outer boundary \citep{AllenR2018}.

\subsubsection{Electrons}
Saturn’s electron radiation belt extends outward of Saturn’s A-ring ($L> 2.27$) and peaks at L$\approx 2.5$ (Fig. \ref{fig:radial}). It does not have a sharp outer boundary at a moon orbit, but decreases with increasing L. Measurable intensities of trapped, MeV electrons typically cease outward of L$\approx 7$ \citep{Roussos2014pss}. The belt's seed population is in the middle magnetosphere (ring current, L$\approx$8-12), from where electrons get accelerated adiabatically at least down to L=4, where high intensities are found up to $\approx 1$MeV. For reference, significant intensities can be found up to several 10 MeV at Jupiter \citep{kollmann2018jgr}.
About 50\% of the seed electrons are field aligned \citep{Clark2014pss}, suggesting that they may be partly maintained through acceleration in the auroral region (Sec. \ref{sec:accel}).

Local acceleration through Whistler mode chorus waves (Sec. \ref{sec:accel}) is expected to be inefficient \citep{Shprits2012jgr} even though the intensity of these waves is significant \citep{Menietti2014}. Radiation belt electrons are mainly lost through scattering (Sec. \ref{sec:scatter}) in neutral material \citep{Lorenzato2012jgr}. Electrons are lost to moons and ring arcs \citep{Hedman2007science,Andriopoulou2012icarus} but the resulting wakes along the electron drift direction are refilled via radial diffusion quicker than the electron-moon re-encounter time \citep{VanAllen1980jgr, Roussos2007jgr}, explaining why electron belts are not segmented as the proton ones at the moon orbits. Long re-encounter times result from electron drifts that are sensitive to global convective flows (Sect. \ref{sec:transport}) and which distort the shape of their nominally circular drift electron orbits, limiting their residence time along the circular sweeping corridors of moons and rings \citep{roussos2016icarus_a}. The same flows are also responsible for persistent local time asymmetries in the spectra of energetic electrons, mostly along the noon-midnight direction \citep{Carbary2009jgr, Paranicas2009jgr, Thomsen2012jgr, Andriopoulou2013icarus}, and the presence of local-time confined belt components (termed ``microbelts'') at the outer edge of Saturn's A-ring \citep{roussos2018grl}.

No energetic electrons were observed close to the planet (L$<$1.22) \citep{Roussos2018science}, consistent with the lack of synchrotron emissions from that region \citep{Griessmeier2011}. Saturn’s main rings block inward transport of electrons coming from beyond the rings to those low L-shells. Local sources like CRAND (Sec. \ref{sec:CRAND}), are not fast enough compared to local electron losses to dust and neutrals (Sec. \ref{sec:friction}, \ref{sec:scatter}).

\subsection{Time variability}

\subsubsection{Ions} 
The most striking change in Saturn's MeV proton belts is their transient extension beyond their nominal outer boundary at L=4.9 \citep{Roussos2008grl}. Such extensions commence when MeV ions of Interplanetary Coronal Mass Ejections (ICME), which penetrate easily down to L$\approx 10$ \citep{Kotovathesis}, get rapidly transported towards L$\approx$5 due to magnetospheric convection induced by the passage of an ICME shock. This process forms transient ion belts which have been observed to last between 3 and 9 weeks. These are the only events known that can induce a fast, global transport of keV ions into low L-shells (L$\approx$4), before charge exchange in the Enceladus neutral cloud depletes them \citep{roussos2018icarus}. Transient, local $<$1 MeV ion flux enhancements from interchange injections (Sec. \ref{sec:transport}) are rarely observed inward of L$\approx$5 and are limited below few hundred keV.

\begin{figure}[htb]
\includegraphics[clip=true,trim=0cm 0cm 0cm 0cm,width=9cm]{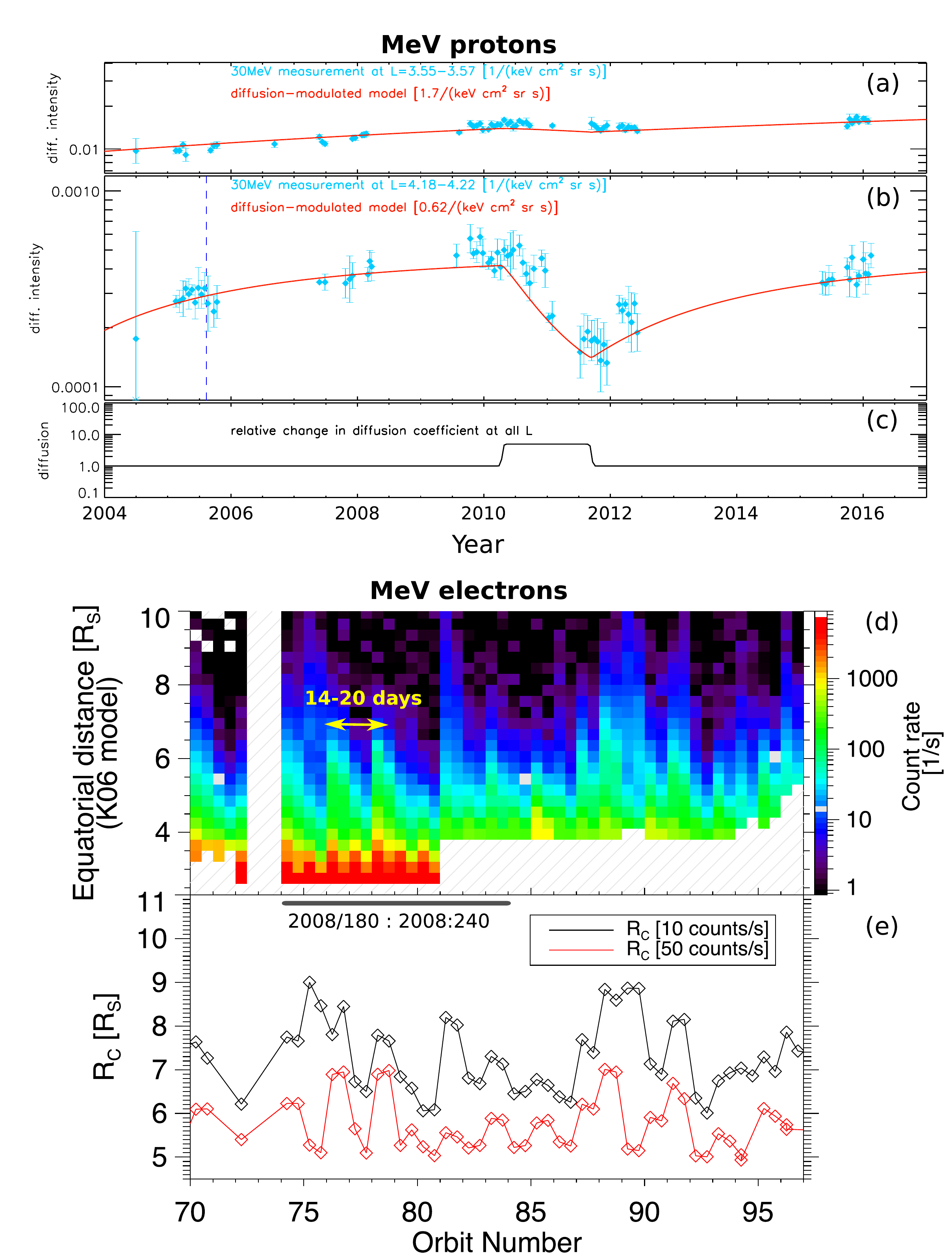}
  \caption{MeV proton and electron variability in Saturn's radiation belts. Panels (a) to (b) show long term changes in $\approx$30 MeV proton fluxes, for two different L-shells. The red curve shows a simulated profile assuming a time-dependent radial diffusion coefficient (c). Panel (d) shows an orbit-distance spectrogram of $>$1 MeV electron count-rates at Saturn, while panel (e) the corresponding electron belt extension, $R_{C}$, for two count-rate thresholds, for a period in 2008. A 14-day periodicity is highlighted in yellow, and a similar period was simultaneously observed in the solar wind (grey bar). Panels (a)-(c) are based on \cite{Kollmann2017nature}, (d)-(e) on \citet{roussos2017icarus}.}%
  \label{fig_saturn_time}
\end{figure}

In comparison, changes in the permanent proton belts are marginal, but have a well-resolved profile (Fig. 3, panels a-c). A slow, steady increase of proton fluxes resulting from the variability of the CRAND source \citep{Roussos2011jgr} is balanced by abrupt flux dropouts over time scales of 1-2 years resulting from variable magnetospheric diffusion modulating the proton loss rates \citep{Kollmann2017nature}.

\subsubsection{Electrons} 

In stark contrast to the proton belts, changes in electron radiation belts are rapid and large in amplitude, affecting their intensity and L-shell profile. The reason is that communication of the electron belts with their highly variable source region in the ring current is not restricted by Saturn's moons \citep{Roussos2014pss}. This variability has multiple sources. Biweekly oscillations of the belts all the way down to A-ring indicate that Corotating Interaction Regions (CIRs) in the heliopshere  influence the electron belts (Fig. 3, panels d-e). Transient electron radiation belts can also follow ICMEs or large scale injections in the magnetotail \citep{roussos2017icarus}. 

Transients typically evolve on time scales faster than 2-3 weeks, much faster than for protons. This difference suggests that electron transport is dominated by variable convection, not stochastic radial diffusion. E.g. CIRs modulate the strength of magnetospheric convection, which in turn controls the radial transport of energetic electrons \citep{roussos2017icarus,roussos2018grl}. Fast. but localized changes in the electron radiation belts have been observed also in the keV energy range. Such electrons penetrate to small L-shells and sustain the radiation primarily through interchange \citep{Paranicas2010jgr}. Tail reconnection can trigger periods of deeply penetrating, recurring interchange carrying keV electrons to low L-shells faster than 10-20 hours \citep{Thomsen2016jgr}. 

Not all aspects of the electron belts' variability have been explained. In particular, a 2-3 month period in 2011, when the electron belts were persistently low in intensity, has no clear explanation. 

\section{Jupiter's radiation belts}\label{sec:Jupiter}

\subsection{Average configuration}

\subsubsection{Ions}
The inner edge of Jupiter’s ion belts is limited by the region of stable magnetic trapping at about $1.2R_J$ \citep{Nenon2018grl}. Jupiter’s ion radiation belts do not have a clear outer boundary but gradually decrease in intensity outward (Fig. \ref{fig:radial}). Different to Saturn, they are less clearly separated by moon orbits. This is because the non-alignment between the magnetic equatorial and the orbital plane of the moons makes ion-moon encounters less likely than at Saturn. This explains why ions of both iogenic and solar wind origin (e.g. helium, oxygen, sulphur at various charge states) exist across the magnetosphere \citep{Cohen2001jgr, Selesnick2009jgr}.

Europa and its gas torus only have a minor impact on the outermost radiation belt \citep{Lagg2003, Kollmann2016europa}. The highest ion intensities at hundreds of keV are observed in this outermost belt that peaks at L$\approx 7$ \citep{mauk2004jgr, kollmann2018jgr}. Ionization of material released from Io and subsequent pickup leads to the generation of EMIC waves, that result in pitch angle diffusion, scattering of ions into the atmosphere \citep{Nenon2018}, and eventually a strong depletion in ion intensities near the moon's L-shell ($L=5.9$) that forms the inner edge of the outer ion belt (\cite{Thomsen1977}, Fig. \ref{fig:radial}). Still, some ions can be transported across Io’s orbit and form another radiation belt peaking at $L\approx 3$ \citep{Garrett2015, Nenon2018}. The next radiation belt segment peaks at $L\approx 2$ and extends until Jupiter’s main ring ($L=1.8$). This is the belt with the highest ion energies ($>350$MeV) \citep{pehlke2013thesis}. The best ion measurements are in helium, suggesting a source from spallation of ring material \citep{Fischer1996science}. Given that the abundance of other species is currently undetermined, contributions from radially transported magnetospheric ions is not excluded \citep{Nenon2018grl}. 
The innermost radiation belt ($L<2$) is populated with such magnetospheric ions having energies of at least 1 MeV. Since outward of that belt intensities of such ions are low, it was suggested that it is sourced by stripped energetic neutral atoms rather than diffusive transport \citep{kollmann2017grl}.

\subsubsection{Electrons} 
Electron intensities around Jupiter are not clearly separated by location, so that it can be said that Jupiter has a single electron radiation belt. The intensities in its core are so high that they are often outside of the specifications that particle radiation measuring instruments were designed for. Therefore there exist relatively few high-quality measurements and inferences of intensities therefore rely on heavily corrected data from particle instruments, scarce measurements from flyby missions, synchrotron emissions, and particle radiation responses of imaging instruments \citep{Taherion2008apj, Soria2016, Becker2017, Nenon2017jgr}.

The seed population of the electron belt is in the outer magnetosphere. Evidence that adiabatic radial transport is a key energy source for MeV electrons down to at least L$\approx$20 has been presented by \citep{kollmann2018jgr}. Seed electrons that can be adiabatically heated are provided by a variety of processes, such as auroral acceleration \citep{Tomas2004jgr}. Whistler mode chorus waves \citep{Menietti2016J} are thought to provide additional acceleration on freshly injected particles \citep{Woodfield2013} but become inefficient after the distributions reach a steady-state \citep{Soria2017}. Pitch angle diffusion limits the electron intensities below the Kennel-Petschek limit, at least around L=8 \citep{Mauk2010jgr}.

Electron intensities drop by a factor of $<10$ at $L\approx 6$, because it is thought that they are absorbed by Io, except at energies that have similar drift speeds as the moon’s orbital speed \citep{santoscosta2001pss}. The absorption losses are enhanced by pitch angle diffusion \citep{Nenon2017jgr}. MeV electron intensities peak at $L\approx 3$ (\cite{Soria2016}, Fig. \ref{fig:radial}). Significant electron intensities at Jupiter exist up to several tens of MeV, as synchrotron emissions indicate and simulations predict \citep{depater1994jgr, Nenon2017jgr}. 

\subsection{Time variability} 

Variability in Jupiter's radiation belts is less explored with in-situ data compared to Saturn, primarily due to the long orbit periods of Galileo and Juno, and the spatial dependencies on planetary SIII longitude \citep{Seidelmann1977} that are difficult to disentangle from L-shell and magnetic local time.   

\subsubsection{Ions} 
\label{jions}
Energetic ions at Jupiter appear much more variable than at Saturn \citep{Paranicas1999jgr, Paranicas2002grl, Mauk1999jgr, Selesnick2001jgr_jupiter_b} but the time scales have not yet been explored. Given that the Galilean moons cannot prevent ion transport, it is expected that variability would develop faster than the year-long changes of Saturn's proton belts and may reflect the rate of radial ion transport and acceleration in the system, estimated to be in the range of weeks to months \citep{Cohen2001jgr}. Transient acceleration events producing MeV ions have been observed to occur at 25-30 $\mathrm{R_{J}}$ \citep{Selesnick2001jgr_jupiter} but their origin is unclear. At lower keV energies, radiation belt ions may be sustained by dipolarization events, which recur every few days, and interchange injections which could be modulated by the aforementioned dipolarizations \citep{Mauk1999jgr,Kronberg2007jgr, Dumont2014jgr, Louarn2014jgr}. The same time scale may then apply for the low-altitude ion belt \citep{kollmann2017grl}, if it is indeed produced by stripping of ENAs generated at the tori of Io and Europa. 

\begin{figure}[t]
\includegraphics[clip=true,trim=0cm 0cm 0cm 0cm,width=13cm]{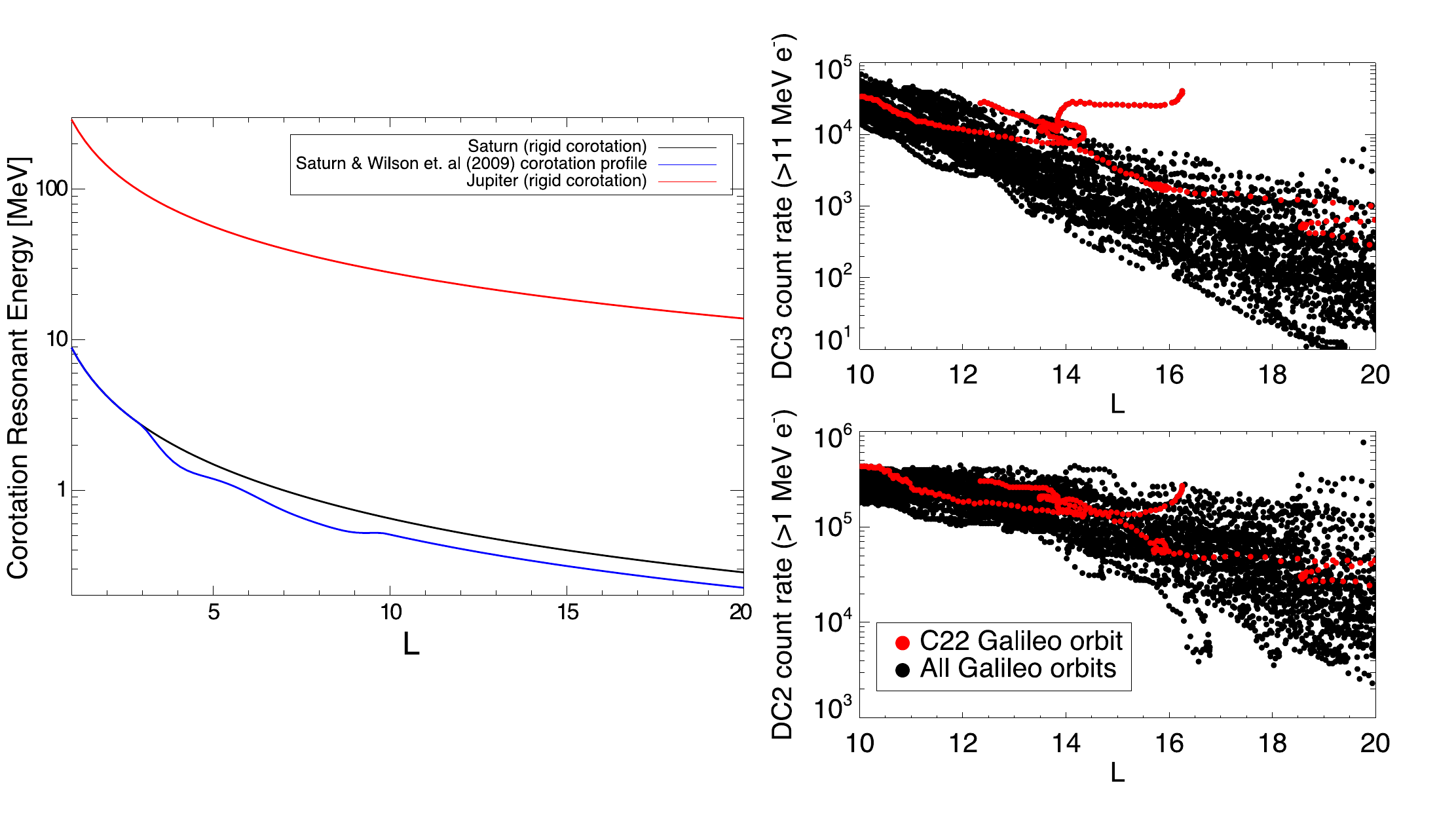}
  \caption{(Left) Electron energies as a function of L-shell, where electron drifts cancel-out with corotation (resonant energies), at Jupiter and Saturn. For Saturn, the blue curve shows these energies for a more realistic profile for corotation based on \cite{Wilson2009grl}. (Right) Galileo EPD measurements of $>$1 MeV and $>$11 MeV electrons as a function of $L$. The unusual C22 orbit shows a strong transient only at $>$11 MeV, close to the resonant energy range for L$\approx$15. From \cite{roussos2018icarus}.}%
  \label{fig_resonant_time}
\end{figure}

\subsubsection{Electrons} 
Variability of the jovian MeV electron belt can be observed remotely through the synchrotron radiation it emits, which reveals that diffusive transport from the electron source region (L$>$6) to its innermost regions lasts about 2 years \citep{depater1994jgr}. Long-term changes in this belt have been positively correlated  to variable solar wind parameters, with time lags of 0.7 or 2.7 years \citep{Santoscosta2008jgr, Galopeau2001pss}. 

Many studies identified additional short or medium-term changes in the synchrotron emissions (e.g. \cite{Miyoshi1999grl}). One of those was linked to the entry of the disrupted comet Shoemaker-Levy 9 (SL9) in Jupiter's magnetosphere and its impact on the planet, with various mechanisms, such as enhanced chorus emissions or shock-induced acceleration, considered to explain the bursts of synchrotron emissions seen at the time (e.g. \cite{depater1995science, bolton1995grl,Brecht2001icarus}). SL9-type impacts are not frequent, but still not rare events, given Jupiter's strong gravity \citep{santocosta2011jgr}. Changes in the solar EUV flux input to the jovian atmosphere correlate with 3-5 days delayed changes in the electron belts \citep{Tsuchiya2011jgr}, which strongly supports the theoretical link between magnetospheric diffusion and variable thermospheric winds (Sec. \ref{sec:transport}).

MeV electron fluxes measured in-situ scatter over one order of magnitude for any given L-shell in the electron belt. More extreme transients, lasting for several days each (Galileo's orbits E06, C22), revealed MeV electron intensities rising above the average by factors of several hundred \citep{russell2001adv,Jun2005Icarus, sorensen2005icarus}. These enhancements were captured only in measurements of ultra-relativistic electrons ($>$11 MeV), hinting a similar acceleration mechanism as at Saturn, where electrons in drift-resonance with corotation are preferentially transported due to variable, convective flows (Fig. 4). 

Such flows may be relevant to a dawn-dusk convective electric field in Jupiter's inner magnetosphere \citep{Barbosa1983grl,Ip1983nature}, the amplitude of which enhances during solar wind induced compressions of the system \citep{murakami2016grl}. When radial diffusion rates estimates take into consideration these dawn-dusk electric field variations, the 0.7 or 2.7 years time lag in the response of the inner electron belts to long-term solar wind changes, can be replicated \citep{Han2018jgr}. A similar electric field but with reversed pointing, controlled by diurnal atmospheric winds, is believed to modulate dawn-dusk asymmetries in the synchrotron part of the belt \citep{Kita2015jgr}.

At lower energies, $<$1 MeV, the intensity scatter between orbits is similarly large and mostly attributed to interchange injections \citep{Mauk1999jgr, Clark2016jgr}.

\section{Summary and conclusions} 

Jupiter's and Saturn's radiation belts are strongly coupled to the characteristics of their host planet. This coupling highlights processes that are unresolved at Earth but may be critical for extraterrestrial radiation belts. Comparative studies between Earth, Jupiter and Saturn are therefore necessary for forming a more universal picture of how radiation belts work. Thanks to the nearly two-decade long datasets from Galileo, Juno and Cassini, such a comparison is now possible.

For instance, a seemingly unimportant detail such as the northward orientation of the jovian and saturnian magnetic fields, which controls the direction of energetic particle magnetic drifts, factors in to explain how very weak convective flows induced by the solar wind have a dominant control over electron transport and acceleration. So, even though Jupiter's and Saturn's magnetospheres are termed as ``corotation-dominated'', their MeV electron belt variability may be solar wind dominated. In addition, the perfect alignment of the magnetic and rotational axes at Saturn amplifies the role of energetic particle losses to moons and rings, explaining the much lower particle fluxes of the kronian belts compared to those of Jupiter, even though the two magnetospheres have many qualitative similarities. The study of the belt's seed populations also turns out to be an important one. Besides large scale dipolarization injections on the nightside magnetosphere, energetic particles accelerated near instantly to MeV energies at the aurora may also become trapped and a subsequent source of the radiation belts.

At present, the Juno mission is operating at Jupiter, collecting high quality, continuous measurements of the planet's magnetosphere and radiation belts at high latitudes and close to planet, regions that have not been covered by the Galileo orbiter. The 13-year Cassini dataset, and in particular the capabilities it offers for multi-instrument radiation belt studies, are gradually getting more exploited. The future Jupiter Icy Moons Explorer (JUICE) and Europa Clipper missions \citep{grasset2013pss, pappalardo2017lpi}, while focusing primarily on the jovian moons Ganymede and Europa, respectively, will carry particles and fields instrumentation, including the first ever ENA imagers to the system, that will further enhance our understanding of the planet's radiation belts. The impact of the expected new measurements and investigations would significantly benefit planning of potential future missions to the magnetospheres of Uranus and Neptune and research on exoplanets which may host radiation belts.


\listofchanges

\end{document}